\documentclass[%
 reprint, 
 amsmath,amssymb,
aps, 
]{revtex4-2}

\usepackage{graphicx}% Include figure files
\usepackage{dcolumn}% Align table columns on decimal point
\usepackage{bm}% bold math
\usepackage{multirow}
\usepackage{xcolor}
\usepackage[normalem]{ulem}

\begin{document}

%\preprint{APS/123-QED}

\title{\textbf{Deciphering the dynamics of nuclear collisions with elongated structure of $^{20}$Ne} 
}% 

\author{Deependra Sharma}
 \email{deep.phy@outlook.com}
\author{Arpit Singh}%
\email{arpit.ehep@gmail.com}
\author{Sadhana Dash}
\email{sadhana@phy.iitb.ac.in}
 
\affiliation{%
 Department of Physics, Indian Institute of Technology Bombay, Mumbai-400076, India
}%

%\date{}% It is always \today, today,

\begin{abstract}
We investigate the role of intrinsic nuclear geometry of $^{20}$Ne nucleus in particle production in small collision systems. Discrete geometrical representations of $^{20}$Ne, including bi-pyramidal $\alpha$-cluster structure in two different configurations along with NLEFT configurations, are implemented within the Monte Carlo Pythia8/Angantyr framework. The resulting particle production observables in $^{20}$Ne-$^{20}$Ne collisions at $\sqrt{s_{NN}}$ = 5.36 TeV are systematically compared with those obtained using conventional Woods-Saxon description as well as with the available hydrodynamic model calculations. We investigate the sensitivity of charged particle multiplicity, transverse momentum distributions and mean transverse momentum $\langle p_T \rangle$ to nuclear geometry, $\alpha$-clustering, and orientation effects of $^{20}$Ne nucleus. While explicit clustering and orientation dependence lead to a noticeable modifications in final state charged particle multiplicity, their impact on transverse momentum spectra and $\langle p_T \rangle$ remain modest in central collisions. The results highlight the role of intrinsic nuclear geometry and specific orientation of the colliding nuclei, providing insight into the dynamics of small systems in non-hydrodynamic particle production framework. 

\end{abstract}

\maketitle
%\tableofcontents

%\section{\label{sec:level1}I\lowercase{ntroduction}}
\section{\label{sec:level1}Introduction}

In recent years, the observation of collective behavior in small collision systems such as 
proton-proton (pp) and proton-nucleus (pA) collisions has challenged the traditional 
understanding of the conditions necessary for the emergence of quark-gluon plasma (QGP) and 
hydrodynamic-like phenomena~\cite{CMS:2012qk,ATLAS:2012cix,CMS:2018loe,CMS:2016fnw,
CMS:2018jrt,PHENIX:2018lia,PHENIX:2013ktj}. The applicability of hydrodynamics to describe the 
collective behavior of the produced particles in small systems is quite challenging and 
has become an intriguing topic of investigation in heavy-ion physics. The initial geometry of 
the collision zone plays a crucial role in the hydrodynamic description of the strongly 
interacting matter produced in high energy nuclear collisions. There are several evidences in 
pp and pA collisions indicating fluctuations in the shape of proton from event to event
~\cite{Eremin:2003qn,PHENIX:2013ehw,TOTEM:2011vxg,McGlinchey:2016ssj}. The inclusion of a fluctuating nucleon substructure in model calculations seemed to improve the mean
transverse momentum and harmonic flow coefficients description for charged and identified particles ~\cite{Mantysaari:2017cni}. The non-spherical nuclear shape either due to quadrupole deformation as seen in $^{238}$U and $^{129}$Xe or subnucleonic structure such as $\alpha$-clusters in $^{12}$C and $^{16}$O introduce anisotropies in the initial energy density profile, even in central collisions in comparison to the spherical nuclei, which possibly leads to collective expansion of the produced matter due to formation of pressure gradients.\par

The effect of $\alpha$-cluster structure in C$-$C and O$-$O collisions on initial and final state observables has been extensively studied via different theoretical models~\cite{Broniowski:2013dia,Bozek:2014cva,Rybczynski:2019adt} with harmonic flow coefficients as a measure sensitive to clusterization. The upcoming O$-$O data from the Large Hadron Collider (LHC) will validate and provide constraints on the theoretical predictions of the $\alpha$-cluster structure of the colliding nuclei. Like $^{16}$O, the $^{20}$Ne nucleus is also a prime candidate to study the influence of $\alpha$-cluster on the initial state geometry in nuclear collisions. The $^{20}$Ne nucleus is known to have a bi-pyramidal bowling-pin like structure composed of five $\alpha$-clusters with four $\alpha$-clusters arranged in a tetrahedral geometry which forms a configuration resembling a $^{16}$O nucleus, and the fifth $\alpha$-cluster is positioned along the symmetry axis, resulting in a $^{16}$O+$\alpha$ composite system. Several theoretical models have predicted the clustering in $^{20}$Ne, including the antisymmetrized quasicluster model (AQCM)~\cite{Yamaguchi:2023mya}, the antisymmetrized molecular dynamics (AMD) ~\cite{Kanada-Enyo:2020goh}, the hybrid-Brink-THSR wave function~\cite{Zhou:2013eca}, the density functional theory (DFT)~\cite{Ebran:2012uv}, with experimental evidences supporting $\alpha$-clustering in $^{20}$Ne ~\cite{Adachi:2020gjp,Nauruzbayev:2017chb}. \par

The presence of an extra $\alpha$-cluster in $^{20}$Ne creates an elongated and asymmetric geometry in comparison to more compactly clustered $^{16}$O nuclei. This elongated structure enables distinct interaction configurations depending upon the orientation of elongated axis of the $^{20}$Ne nuclei relative to the beam direction. The various possible configurations are Tip-Tip (TT), Body-Body (BB), Body-Tip (BT), and minimum-bias random orientation where all the interaction configurations are equally likely. In Tip-Tip (TT) collisions, in general, the long axes of the colliding nuclei are aligned along the beam direction resulting in a more compact and circular overlap region. In Body-Body (BB) collisions, the long axes of the colliding nuclei are aligned perpendicular to the beam direction, resulting in a highly elliptical overlap region. In Body-Tip (BT) collisions, the long axis of one nucleus is aligned along the beam direction while the other nucleus is oriented perpendicular to the beam direction, resulting in an asymmetric overlap region.\par

\begin{figure*}
\includegraphics[scale=0.7]{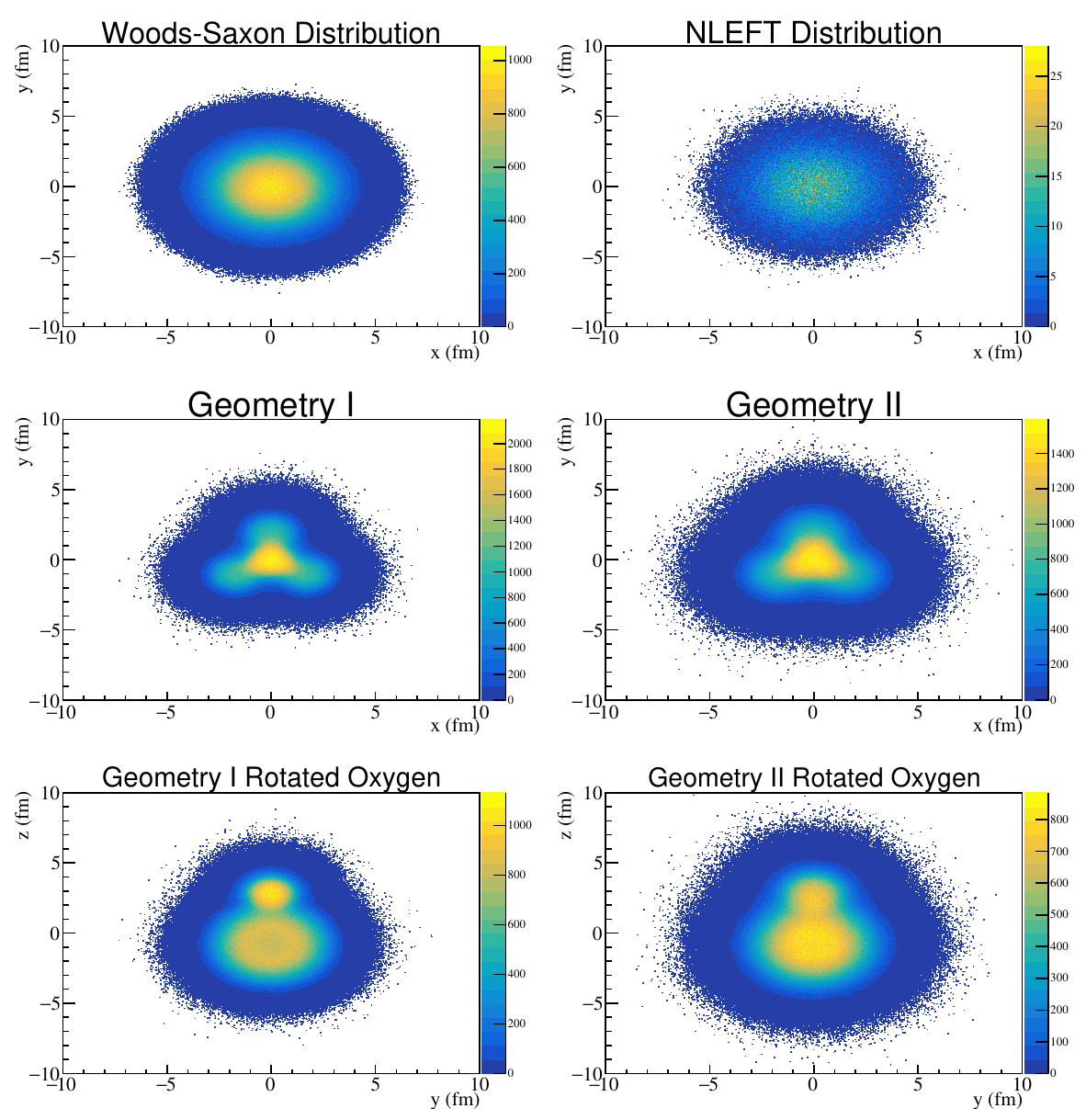}% Here is how to import EPS art
\caption{\label{fig:1} Distribution of nucleons for $^{20}$Ne nuclei obtained using Pythia/Angantyr for Woods-Saxon distribution (top left), NLEFT distribution (top right), Geometry I+
(middle left), and Geometry II (middle right) in x-y plane. The bottom panels display the nucleon distribution of $^{20}$Ne under rotation about the y-axis for Geometry I and II, respectively. The color scale represents the nucleon density in the transverse plane.}
\end{figure*}

These type of configurations has been investigated previously in collisions involving prolate shaped nuclei such as $^{238}$U and $^{129}$Xe, highlighting their impact on different final state observables~\cite{Nepali:2007an,Haque:2011aa,Bairathi:2015uba,Luo:2023syp}. In these heavy nuclei, the deformation arises from the quadrupole deformation of the nuclear shape which leads to continuous and smooth spatial anisotropies. In contrast, $\alpha$-clustering in $^{20}$Ne introduces localized density hot-spots with discrete geometry arrangement, resulting in a more granular and fluctuating initial conditions. The presence of localized density fluctuations arising from alpha clustering can modify the initial eccentricities associated with different collision configurations of $^{20}$Ne, potentially leading to geometric features that differ from those observed in collisions involving deformed heavy ions with smooth nuclear density distributions. \par

In this article, we investigate the effect of the intrinsic structure of the $^{20}$Ne nucleus on the dynamics of small collision systems. We implement discrete geometrical representations of the 
$^{20}$Ne nucleus, including bi-pyramidal $\alpha$-cluster configurations and positively weighted nuclear lattice effective field theory (NLEFT) configurations, within the Monte Carlo Pythia8/Angantyr framework. In addition, we study the effect of different orientations of the elongated $^{20}$Ne nucleus on particle production in $^{20}$Ne$-$$^{20}$Ne collisions at $\sqrt{s_{NN}}$ = 5.36 TeV. Results obtained with the clustered geometries are systematically compared with those from a conventional Woods–Saxon description of the nucleus, as well as with available hydrodynamic model calculations. Focusing on final state particle production in the absence of hydrodynamic evolution, the study isolates the role of intrinsic nuclear geometry and clustering effects, allowing a direct comparison with the medium induced effects present in hydrodynamic descriptions. The remainder of the article is organized as follows. Section II briefly describes the PYTHIA8/Angantyr model and the implementation of the $\alpha$-cluster structure in the $^{20}$Ne nucleus. Section III presents the results and discussion, and Section IV summarizes the main conclusions.

%We study $^{20}$Ne-$^{20}$Ne collisions at $\sqrt{s_{NN}}$ = 5.36 TeV and examine how differences in nuclear geometry and clustering influence particle production observables in the absence of hydrodynamic evolution. This approach allows us to isolate the role of intrinsic nuclear geometry and clustering effects on final-state particle production and to contrast them with medium-driven effects present in hydrodynamic descriptions. The remainder of the article is organized as follows. Section II briefly describes the PYTHIA8/Angantyr model and the implementation of the $\alpha$-cluster structure in the $^{20}$Ne nucleus. Section III presents the results and discussion, and Section IV summarizes the main conclusions.

\section{Collision Framework}

Pythia8/Angantyr~\cite{Bierlich:2018xfw} model is an extension of the Pythia8 event generator, designed to simulate high-energy heavy-ion collisions by incorporating a Glauber-based approach of nuclear geometry and interactions~\cite{Bialas:1976ed}. It models heavy-ion events as a superposition of independent but fluctuating nucleon-nucleon (NN) collisions, incorporating nucleon-level fluctuations as described by Glauber-Gribov theory~\cite{Gribov:1968jf,Heiselberg:1991is}. A central feature of Angantyr is its incorporation of multi-parton interactions (MPI), which are essential in capturing the rich structure of the underlying event. MPI accounts for multiple partonic scatterings within each NN collision, contributing significantly to particle production and energy density, while final-state hadronization is handled by the Lund string model~\cite{Andersson_2023}. The MPI framework is extended to effectively handle the overlapping and correlated interactions in nuclear collisions, allowing for a consistent treatment of soft and semi-hard QCD dynamics across different collision systems. The parameters of the different components of the Angantyr model, such as Glauber–Gribov color fluctuations, MPI, hadronization, etc., are tuned to data from smaller systems ($e^{+}e^{-}$,$ep$, and $pp$) rather than to heavy-ion data.\par

To incorporate the $\alpha$-cluster structure of $^{20}$Ne in a bi-pyramidal geometry, an oxygen nucleus with intrinsic $\alpha$-clustering was first constructed following the approach outlined in  Ref.~\cite{Sharma:2025jhs}. In this configuration, all four $\alpha$-clusters are equidistantly arranged with an inter-cluster spacing of 3.42 fm. Three clusters lie in the x-y plane, while the fourth is positioned along the negative z-axis. The $^{20}$Ne nucleus is subsequently formed by placing a fifth $\alpha$-cluster along the positive z-axis at a distance of 3.6 fm from the center of the oxygen nucleus~\cite{Zhou:2013eca}. The position of the nucleons within the $\alpha$-clusters is sampled from the Woods-Saxon distribution represented using a three-parameter (3pF) Fermi distribution which is given by:
\begin{equation}
\rho(r) = \frac{\rho_0(1+w\left(\frac{r}{R}\right)^2{})}{1 + \exp\left(\frac{r - R}{a}\right)} ,
\end{equation}
where $\rho_0$ is the normalization constant, $R$ is the radius of the nucleus, $a$ is the surface thickness parameter, and $w$ is the skin thickness parameter. Following this prescription, two geometries of the $^{20}$Ne nucleus are constructed, referred to as Geometry I and Geometry II. In Geometry I, the nucleons within each $\alpha$-cluster are sampled using a Woods–Saxon distribution with parameters $R$=0.964 fm, $a$=0.322 fm, and w=0.517~\cite{Li:2020vrg}. With this choice of parameters, an rms radius of 2.76$\pm$0.16 fm is obtained for the $^{20}$Ne nucleus, which is slightly smaller than the experimental value of 
3.0055±0.0021 fm~\cite{Angeli:2013epw}. In order to achieve an experimentally observed nuclear radius, we optimized the Woods-Saxon parameters for the $\alpha$-clusters. This we achieved through minimization of a logarithmic loss function keeping the $^{20}$Ne configuration same as Geometry I and constraining the rms radius of $^{20}$Ne at 3.00 fm while varying the Woods-Saxon parameters over the ranges $R$$\in$[0.8 fm, 1.13 fm], $a$$\in$[0.2 fm, 0.42 fm] and $w$$\in$[0.3, 0.7] for $^{4}$He nucleus with a grid spacing of 0.01 for each parameter. From the grid search, the optimal parameter values were found to be $R$ = 1.12 fm, $a$ = 0.44 fm and $w$ = 0.30. With these parameters, an rms radius of 3.00 $\pm$ 0.20 fm is obtained for $^{20}$Ne nucleus using Angantyr model calculations which is comparable with the experimental values. This configuration is referred to as Geometry II.  \par
In addition to the two geometries described above, positively weighted NLEFT configurations for the $^{20}$Ne nucleus are also employed in the Angantyr model. These configurations, are taken from Ref.~\cite{Giacalone:2024luz}, contain explicit cluster information and comprise of a total of 13589 independent $^{20}$Ne configurations. These configurations result in an rms radius of 3.08$\pm$0.28 fm for the $^{20}$Ne nucleus. In order to isolate the effect of clustering, we have also simulated $^{20}$Ne collisions within the Angantyr model by employing the Woods-Saxon nuclear charge density distribution for the $^{20}$Ne nucleus as described by Eq. 1. The Woods-Saxon distribution parameters used are $R$ = 2.791 fm, $a$ = 0.698 fm, and $w$ =  -0.168~\cite{DeVries:1987atn}, resulting in an rms radius of 3.00 $\pm$ 0.23 fm.

\begin{figure}[h]
\includegraphics[width=\columnwidth]{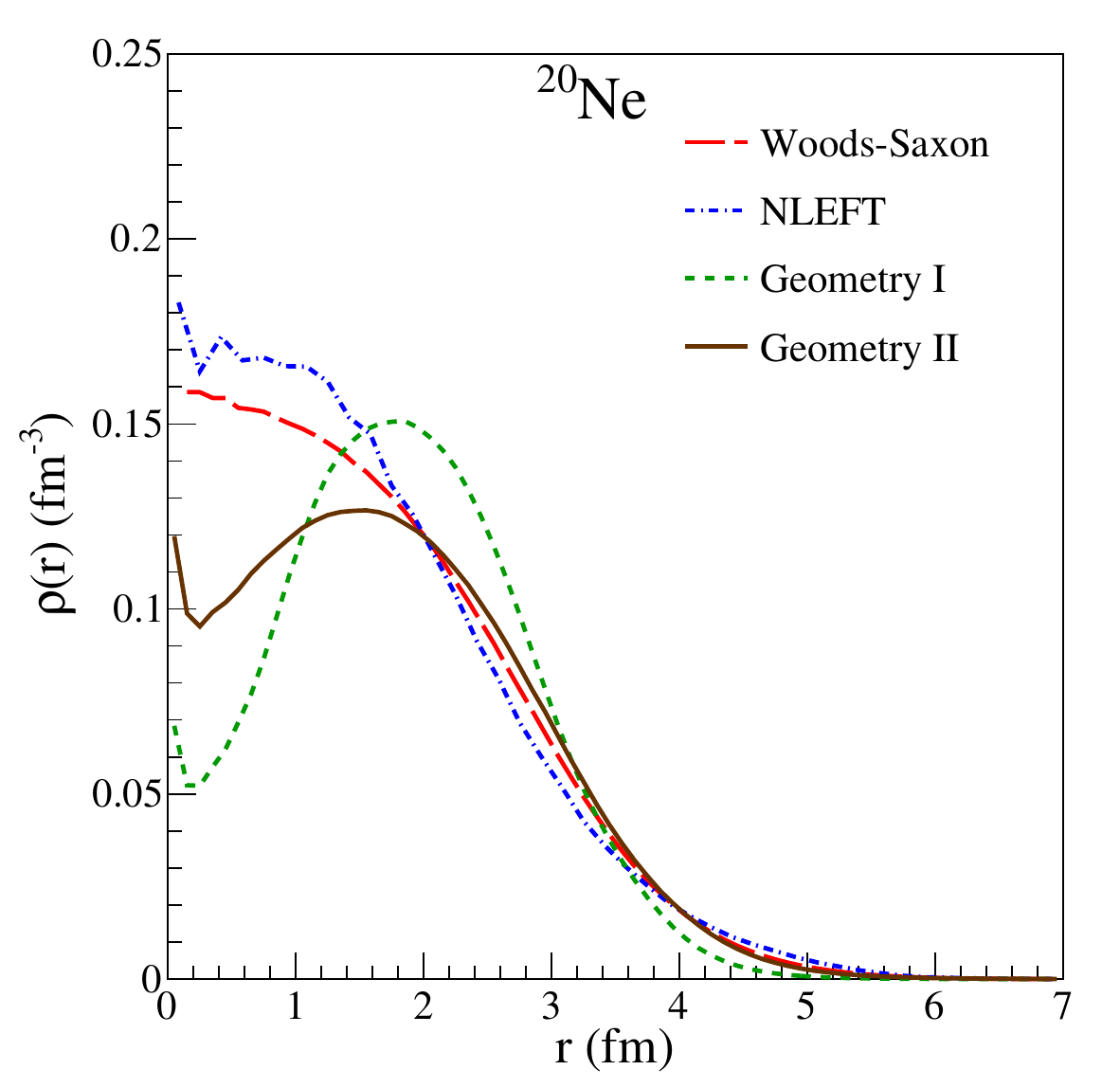}% Here is how to import EPS art
\caption{\label{fig:2} Nuclear charge density distribution for Woods-Saxon, NLEFT, Geometry I and II of the $^{20}$Ne nucleus.}
\end{figure}

\begin{figure*}
\includegraphics[scale=0.7]{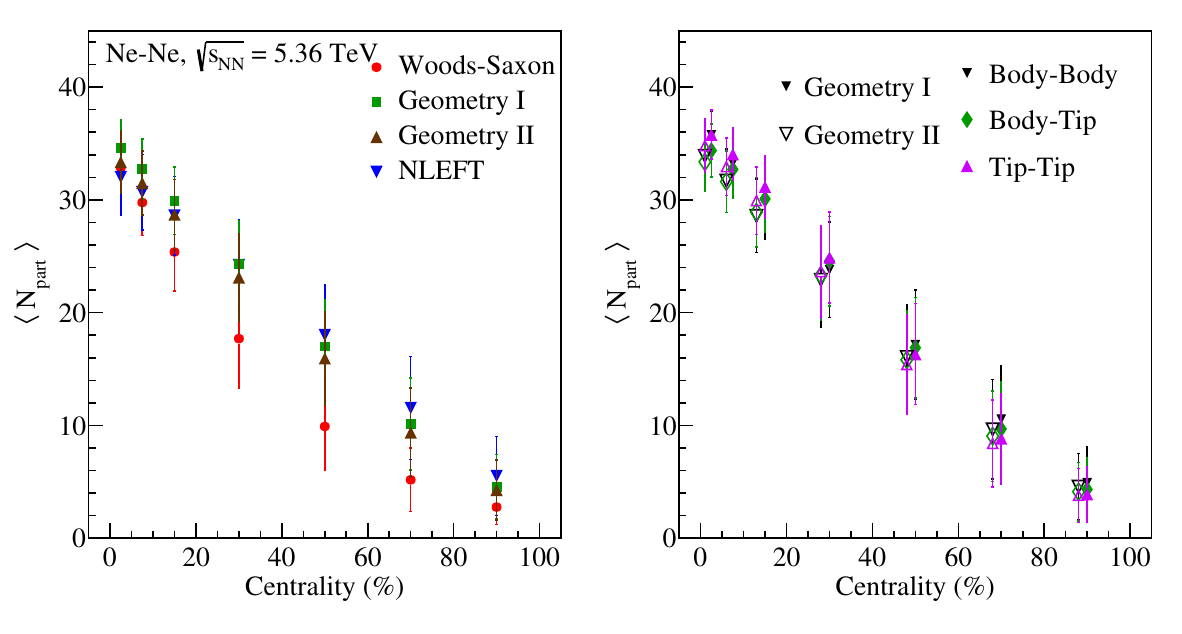}% Here is how to import EPS art
\caption{\label{fig:3} (Left panel) Variation of average number of participants ($\langle N_{part}\rangle$) as a function of centrality (\%) class for Woods-Saxon distribution, NLEFT, and Geometry I and II of the $^{20}$Ne nucleus. (Right panel) Same as left panel, but shown separately for Body-Body, Body-Tip and Tip-Tip orientations of Geometry I and II. For improved visibility, the Geometry II results are shown with open markers and are slightly shifted to the left.} 
\end{figure*}

The two-dimensional density profiles of $^{20}$Ne obtained for the transverse plane of the collision are shown in top and middle panels of Fig~\ref{fig:1} for Woods-Saxon distribution, NLEFT, Geometry I and II. The bottom panels of Fig.~\ref{fig:1} represent density distribution of $^{20}$Ne for Geometry I and Geometry II in y-z plane by applying Euler rotation to oxygen  nucleus only. Both the geometries clearly reveals the formation of bowling-pin like structure of $^{20}$Ne, although the actual simulation is performed by applying three consecutive Euler rotations to the $^{20}$Ne nucleus on an event-by-event basis to randomize the nucleon positions. Fig.~\ref{fig:2} shows the nuclear charge density distribution of $^{20}$Ne as a function of radial distance from the center of the nucleus for Woods-Saxon distribution, NLEFT, Geometry I and II. One peculiar observation for the considered cluster structures is the maximum charge density at $r\sim0$ which then drops suddenly with slight increase in $r$. This behavior arises from the spatial arrangement of the $\alpha$-clusters in the bi-pyramidal geometry, where the clusters are located near the edges of the configuration, leading to a higher probability of finding nucleons away from the geometric center. Consequently, the intrinsic density near the center of the nucleus is strongly suppressed. However, the presence of the fifth $\alpha$-cluster introduces a slight elongation of the $^{20}$Ne geometry, resulting in a small displacement of the centroid from the origin prior to re-centering. Upon event-by-event re-centering, this asymmetry leads to a modest enhancement of the radial charge density at $r\sim0$, followed by a local minimum and a subsequent increase to a maximum, after which the density exhibits the expected exponential decreasing characteristic of the Woods–Saxon distribution. The nuclear charge density distribution, also middle and bottom panels in Fig.~\ref{fig:1}, suggest that the nucleons in Geometry II are more spatially diffuse in comparison to the Geometry I. For the NLEFT configuration, the nuclear charge density is found to be systematically higher than than that obtained from Woods-Saxon parameterization for $r\leq2$ fm. Nevertheless, the overall radial dependence remains qualitatively similar to the Woods-Saxon distribution.\par

Fig.~\ref{fig:3} shows the centrality dependence of average number of participants, $\langle N_{part} \rangle$, in $^{20}$Ne-$^{20}$Ne collisions at $\sqrt{s_{NN}}$ = 5.36 TeV. Left panel shows the variation of $\langle N_{part} \rangle$ for Woods-Saxon distribution, NLEFT, and Geometry I and II configurations obtained from Angantyr model, while right panel represents the corresponding results for BB, BT and TT orientations of $^{20}$Ne nucleus in Geometry I and Geometry II. The observed variations in $\langle N_{part} \rangle$ primarily reflect differences in the geometric overlap region and, consequently, in the number of effective nucleon–nucleon interactions for different configurations and orientations in $^{20}$Ne-$^{20}$Ne collisions, highlighting the role of $\alpha$-cluster structure in modifying the underlying geometrical parameters. 

\section{Results and Discussions}

\begin{figure}
\includegraphics[width=\columnwidth]{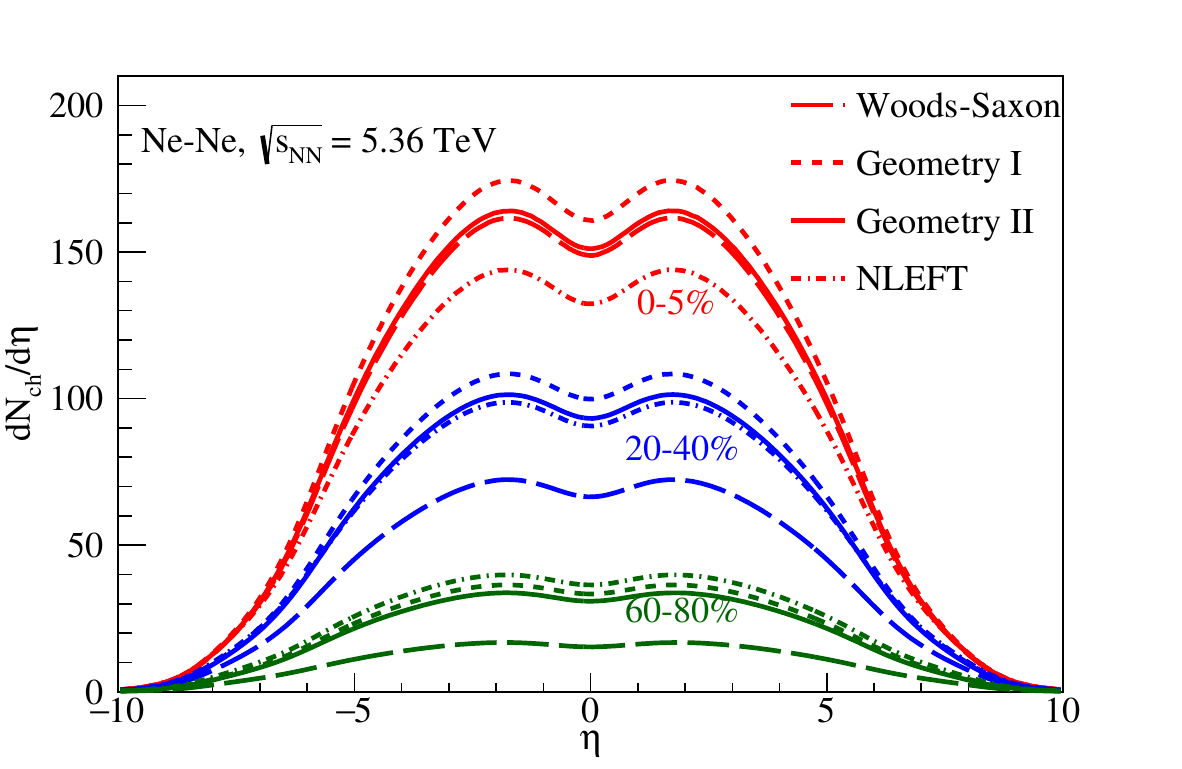}% Here is how to import EPS art
\caption{\label{fig:4} Pseudorapidity distribution of charged hadrons in $^{20}$Ne$-$$^{20}$Ne collisions at $\sqrt{s_{NN}}$ = 5.36 TeV for Woods-Saxon, NLEFT and bi-pyramidal geometries of $^{20}$Ne in (a) 0-5\%, (b) 20-40\%, and (c) 60-80\% centrality classes.}
\end{figure}

\begin{figure}
\includegraphics[width=\columnwidth]{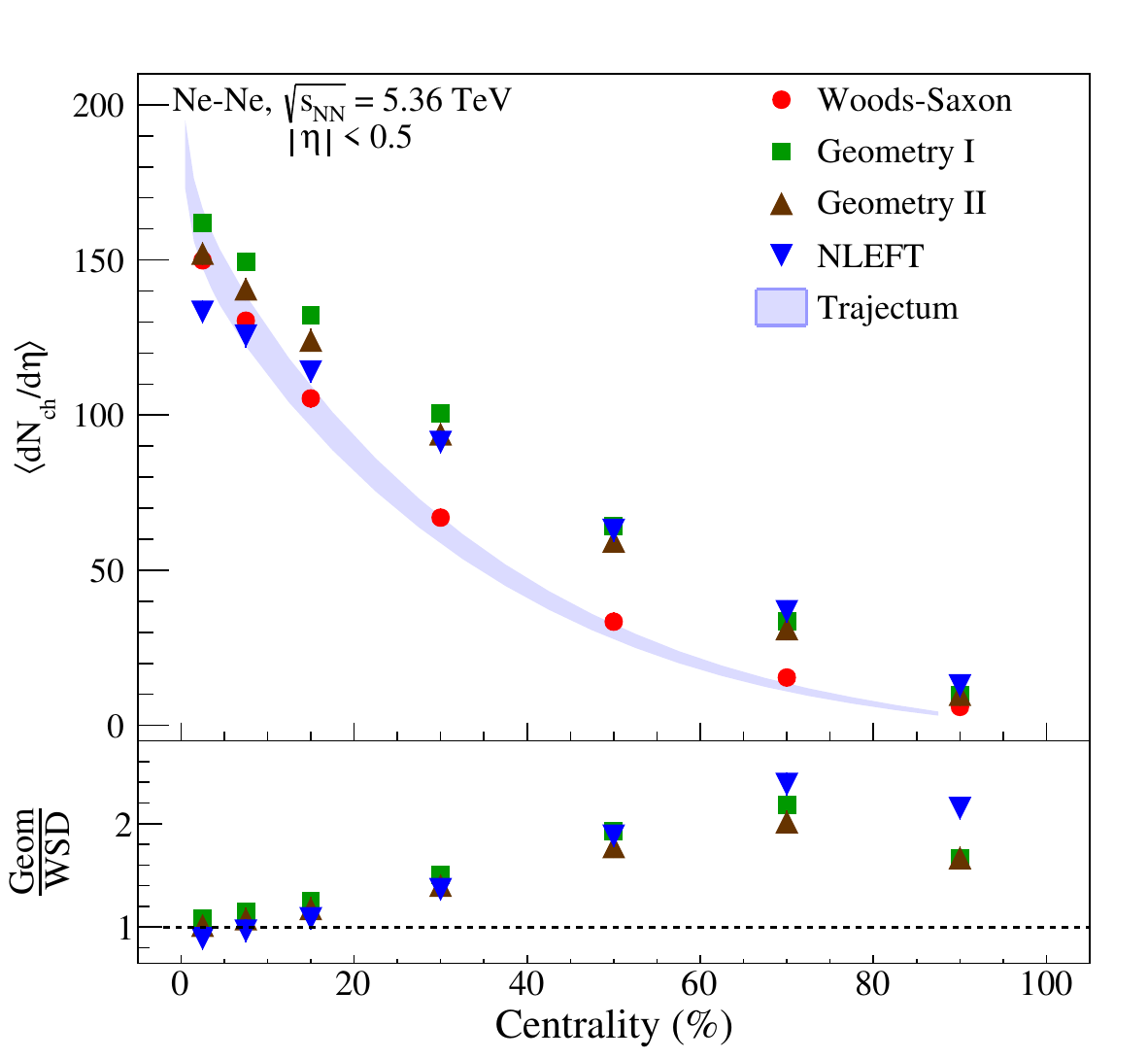}% Here is how to import EPS art
\caption{\label{fig:5} (Top panel) Mean multiplicity of charged hadrons at mid-rapidity in $^{20}$Ne$-$$^{20}$Ne collisions at $\sqrt{s_{NN}}$ = 5.36 TeV for Woods-Saxon, NLEFT and bi-pyramidal geometries. Results from Trajectum model calculations is also shown in the top panel. Bottom panel shows the ratios of mean multiplicities of Geometry I, II and NLEFT to Woods-Saxon distribution a a function of centrality.}
\end{figure}

The results obtained for the pseudorapidity distribution of charged hadrons in $^{20}$Ne$-$$^{20}$Ne collisions at $\sqrt{s_{NN}}$ = 5.36 TeV using the Pythia8/Angantyr model are displayed in Fig.~\ref{fig:4}. The pseudorapidity distribution is a crucial observable that provides insights into the particle production dynamics and the initial state geometry of the collision. The distribution is shown for the 0-5\%, 20-40\%, and 60-80\% centrality classes for Woods-Saxon, NLEFT, and Geometry I and II. In all the three centrality intervals, a clear distinction between the charged hadron production is visible for different $^{20}$Ne configurations, which remain visible even at larger rapidities. Fig.~\ref{fig:5} shows the mean charged particle multiplicity at mid-rapidity obtained for $^{20}$Ne-$^{20}$Ne collisions at $\sqrt{s_{NN}}$ = 5.36 TeV. In the top panel, the results are shown for Woods-Saxon, NLEFT, and Geometry I and II obtained from Angantyr model. In the bottom panel, the corresponding ratios of NLEFT, Geometry I and II to those obtained from Woods-Saxon distribution is displayed. Results from the published Trajectum model employing NLEFT initial conditions for $^{20}$Ne nucleus are also shown for comparison. \par

In the Trajectum model, nucleon configurations obtained from Nuclear Lattice Effective Field Theory (NLEFT) are used to initialize the colliding nuclei~\cite{Giacalone:2024luz}. For each event, participant nucleons are determined and the initial energy density is deposited in the transverse plane using a Trento-type ansatz, followed by a short pre-equilibrium stage and relativistic viscous hydrodynamic evolution until hadronization. The subsequent hadronic rescattering and decays are simulated using the SMASH transport model. The model parameters are selected probabilistically by sampling from the posterior distribution obtained in a Bayesian analysis of Pb-Pb collisions~\cite{Giacalone:2023cet}. The uncertainties arising from the parameter variations are estimated by sampling twenty parameter sets randomly drawn from the posterior distribution. \par

In Angantyr model calculations, a clear separation in charged particle multiplicity is observed for all the configurations with explicit implementation of $\alpha$-cluster structures in central collisions which reduce progressively and tend to merge together towards peripheral collisions. This can be understood in terms of the effective nuclear size and spatial distribution of nucleons. A more compact nuclear configuration leads to a smaller effective interaction area. resulting in a higher number of binary nucleon-nucleon interactions per unit overlap volume. This. in turn, increases the string density and hence an enhances the charged particle multiplicity. The effective size of the $^{20}$Ne nucleus follows the ordering NLEFT $>$ Geometry II $>$ Geometry I. Correspondingly, the charged particle multiplicity exhibits the inverse hierarchy, Geometry I $>$ Geometry II $>$ NLEFT. As the collisions become non-central, the effective overlap region decreases leading to a reduced number of nucleon-nucleon interactions, making final state charged particle multiplicity mostly insensitive to the underlying nuclear geometry. \par

A notable  observation is that the charged particle multiplicity obtained with the Angantyr model using Woods-Saxon distribution, without $\alpha$-cluster structure, is consistent with the corresponding results from the Trajectum model calculations. In contrast, configurations with explicit cluster structure implemented in the Angantyr model exhibit clear deviation from the Woods-Saxon and Trajectum model results. The NLEFT configuration used in the Angantyr model follows a trend similar to that observed for Geometry I and Geometry II. Angantyr model translates initial nuclear geometry into final state particle multiplicity through nucleon-nucleon interactions. The presence of explicit clustering enhances the local overlap density of strings, leading to a significant increase in charged particle multiplicity in mid-central and peripheral collisions in comparison with Woods-Saxon distribution having a smooth nuclear density profile. The observed differences between Trajectum and Angantyr results with explicit cluster structures can be attributed to the distinct particle production mechanism of the two models. In hydrodynamic models like Trajectum, the final state charged particle multiplicity is primarily governed by the total initial energy or entropy density deposited in the system. The hydrodynamic evolution converts energy density profile into final state hadrons through collective expansion and freezeout, effectively smoothing small scale fluctuations. Consequently, Trajectum yields mean multiplicities comparable to the Woods-Saxon geometry obtained using Angantyr model. \par

The apparent agreement between the Woods–Saxon case in Angantyr and the corresponding predictions from Trajectum should therefore be interpreted with caution and does not by itself establish that the Woods–Saxon charge density distribution provides the correct description of the initial nuclear geometry. In the absence of experimental measurements for $^{20}$Ne-$^{20}$Ne at $\sqrt{s_{NN}}$ = 5.36 TeV collisions, the results from Trajectum represent model predictions obtained by sampling parameters within a reasonable range. Therefore, the results presented only illustrates that the Woods–Saxon scenario lies within the range of Trajectum model predictions, rather than identifying it as the preferred initial-state description for the $^{20}$Ne nucleus.

\begin{figure}[h]
\includegraphics[width=\columnwidth]{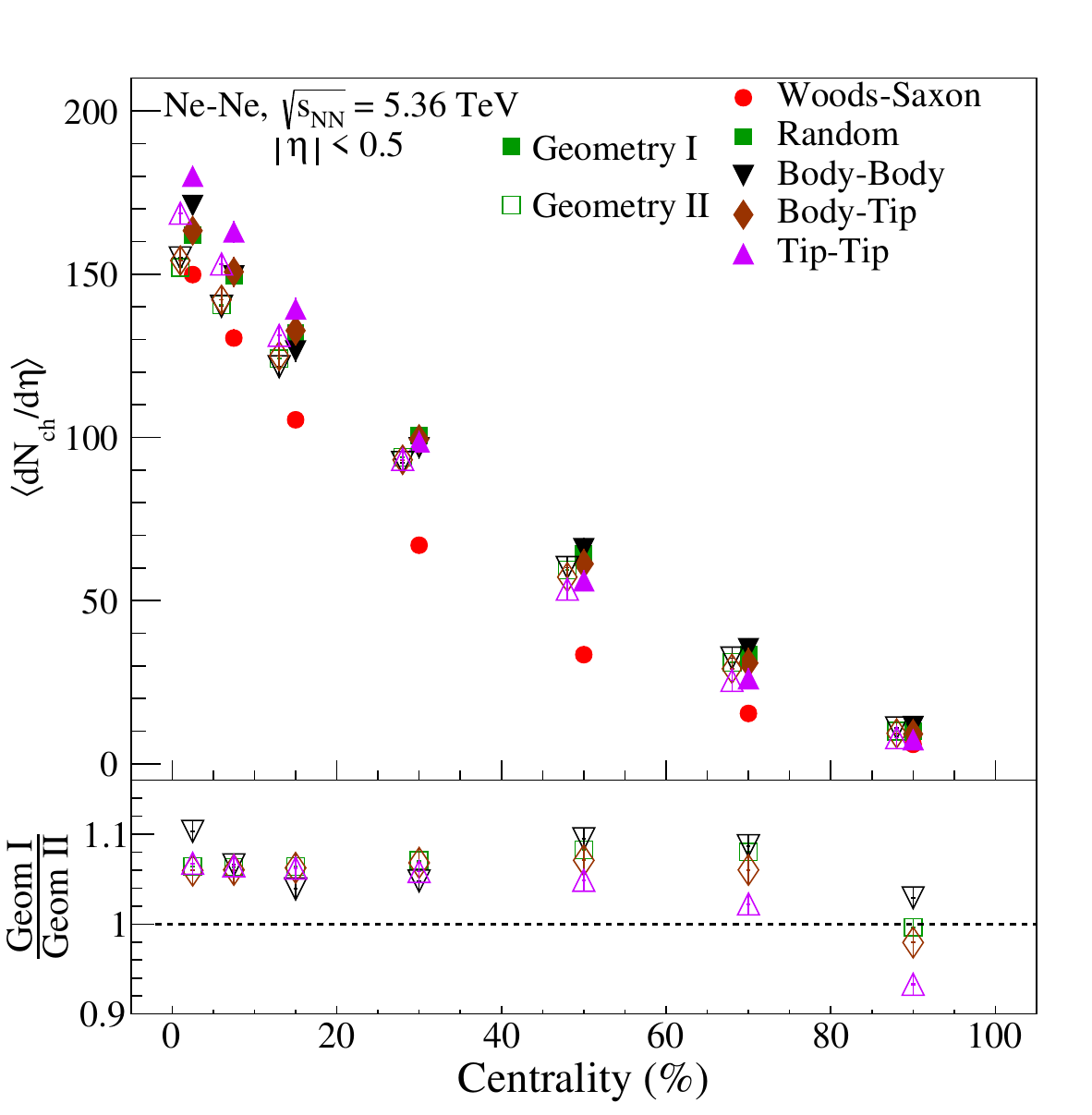}% Here is how to import EPS art
\caption{\label{fig:6} (Top panel) Mean multiplicity of charged hadrons for Geometry I and Geometry II in $^{20}$Ne-$^{20}$Ne collisions for Body-Body, Body-Tip and Tip-Tip orientations of $^{20}$Ne nucleus. For better visibility, the Geometry II results are shown with open markers and are slightly shifted to the left. Bottom panel presents the Geometry I to Geometry II ratio for the corresponding nuclear orientations.}    
\end{figure}

Once the orientation of the colliding nuclei is fixed, the overlap area changes, which affects the number of participating nucleons and leads to corresponding variations in the charged-particle multiplicity. Fig.~\ref{fig:6} shows the centrality dependence of charged particle multiplicity at mid-rapidity for Geometry I and Geometry II in BB, BT, TT and random orientations along with Woods-Saxon distribution in $^{20}$Ne-$^{20}$Ne collisions at $\sqrt{s_{NN}}$ = 5.36 TeV. The Geometry I to Geometry II ratios for the corresponding orientations are also shown in the bottom panel. Among the various orientations, in central collisions, TT collisions yield the highest multiplicity, followed by BB collisions, while BT and randomly oriented collisions results in comparatively lower and similar multiplicities. In TT collisions, the highest overlap density is generated in the transverse plane because of the alignment of long axes of both the nuclei with beam causing a large number of nucleon-nucleon interactions in comparison to BB collisions where overlap area is larger but a smaller local density in the transverse plane, resulting in fewer nucleon-nucleon interactions and hence smaller charged particle density than TT collisions. BT collisions have a mismatched overlap geometry which reduces the number of effective interacting nucleons leading to a lower multiplicity than the BB collisions. Random orientations represent an orientation averaged overlap where high density orientations like TT or even BB are statistically rare, resulting in charged particle multiplicity comparable to BT collisions. \par

As collisions become non-central, the overlap region decreases, leading to misalignment of $\alpha$-clusters across different orientations. This weakens the geometrical advantages seen in central collisions, causing significant changes in multiplicity among fixed orientations. Despite of this, the mean multiplicity of charged particles remains comparable for BT and random orientations for all the centrality classes. The multiplicity in TT collisions gradually approaches to random orientation results with increasing centrality, becomes comparable in the 20-40\% centrality interval, and falls below the random orientation results in more peripheral collisions. With increasing centrality, the overlap region in the transverse plane for TT collisions is progressively reduced, corresponding to a narrower effective impact parameter range compared to random orientations. The shrinking overlap region in TT collisions further limits the alignment of the finite size $\alpha$-cluster structures, reducing the number of effective nucleon–nucleon interactions. As a result, the charged particle multiplicities become comparable in mid-central collisions and falls below it in peripheral collisions. \par

A striking change in the behavior of the multiplicity density is observed in BB collisions. As the collisions evolve from central to mid-central, the multiplicity density in BB decreases and becomes lower than all other orientations. The reduction in particle multiplicity can be attributed to the shrinking of transverse overlap region in slightly off-central collisions, which causes some of the high density $\alpha$-clusters to lie outside the interaction zone, while misalignment among the remaining $\alpha$-clusters in the resulting overlap region further reduces the effective number of interacting nucleons. In contrast, TT or BT orientations are characterized by a more compact transverse overlap region which allows a larger fraction of $\alpha$-clusters to participate in the interaction, resulting in a comparatively higher multiplicities. Interestingly, as the collisions evolve from mid-central to peripheral, the multiplicity density in BB collisions becomes highest among all the orientations. This behaviour can be explained by the broader spatial distribution of $\alpha$-clusters in the transverse region because of which some of the clusters can still interact effectively compared to short transverse overlap of TT and BT configurations. Consequently, the charged particle multiplicity in BB collisions increases in the peripheral collisions and eventually exceeds that observed for the other orientations. All the effects are qualitatively similar for the different orientations in Geometry I and Geometry II. The only difference is that the charged particle yields are slightly higher in Geometry I compared to Geometry II,  illustrated by the Geom I to Geom II ratio plot in the bottom panel of Fig.~\ref{fig:6}.

\begin{figure}[h]
\includegraphics[width=\columnwidth]{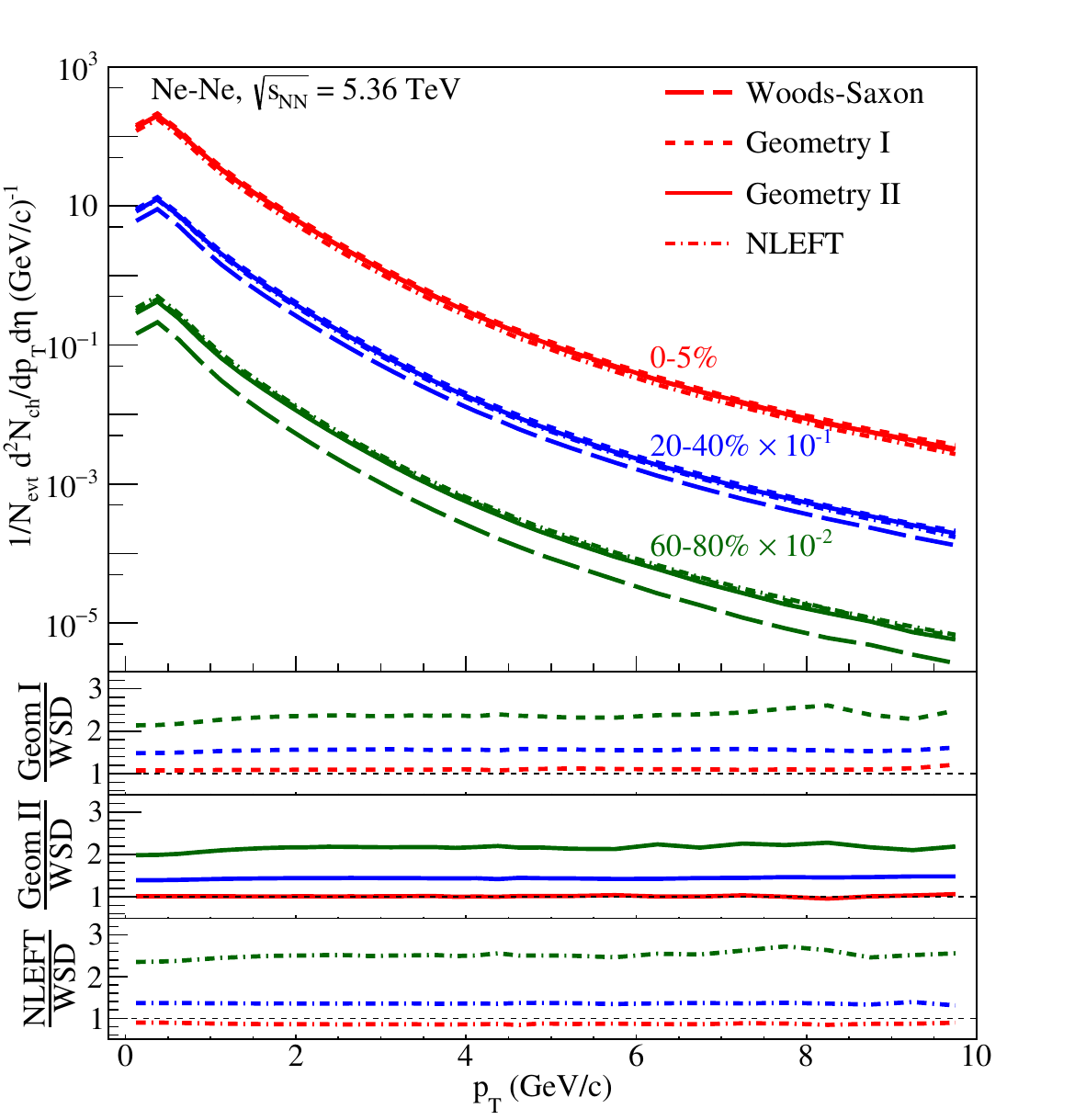}% Here is how to import EPS art
\caption{\label{fig:7} (Top panel) Transverse momentum ($p_T$) distribution of charged hadrons in $^{20}$Ne$-$$^{20}$Ne collisions at $\sqrt{s_{NN}}$ = 5.36 TeV for Woods-Saxon, NLEFT and bi-pyramidal geometries in 0-5\%, 20-40\%, and 60-80\% centrality classes. Bottom panels present the ratios of NLEFT and Geometry I and II to the Woods–Saxon spectra.}    
\end{figure}

Fig.~\ref{fig:7} shows the transverse momentum ($p_T$) distribution of charged hadrons produced in $^{20}$Ne-$^{20}$Ne collisions at $\sqrt{s_{NN}}$ = 5.36 TeV. The results are shown for Woods-Saxon distribution, NLEFT, and Geometry I and II in three centrality intervals. The bi-pyramidal geometries to the Woods-Saxon ratios are shown in the bottom panels. The $p_T$ distribution does not show any significant difference between Woods-Saxon and bi-pyramidal geometries of $^{20}$Ne nucleus in central collisions. However, in mid-central collisions and peripheral collisions, a small enhancement in terms of particle yield is observed for bi-pyramidal geometries along with a slight hardening of the $p_T$ spectra in peripheral collisions. It indicates that the $\alpha$-cluster structure of $^{20}$Ne does not significantly modify the shape of the $p_T$ spectra, it primarily affects the overall particle yield. The Angantyr model describes nuclear effects without hydrodynamics evolution, relying heavily on multi-parton interactions (MPI) and the Lund string model for hadronization. The low-$p_T$ region which is dominated by soft interactions and mid-$p_{T}$ region influenced by semi-hard processes such as minijets are modeled through MPIs arising from independent nucleon-nucleon collisions. The particle yield in these regions results from the superposition of many such independent interactions. High-$p_T$ region is dominated by hard processes which are modeled using perturbative QCD (pQCD) matrix elements. consequently, the slight variation observed in the particle yield for Woods-Saxon and bi-pyramidal geometries is, therefore, due to the number of effective nucelon-nucleon collisions and the local density of partons or strings which are influenced by the spatial alignment of the $\alpha$-clusters for various configurations during the collisions. \par 

\begin{figure}[h]
\includegraphics[width=\columnwidth]{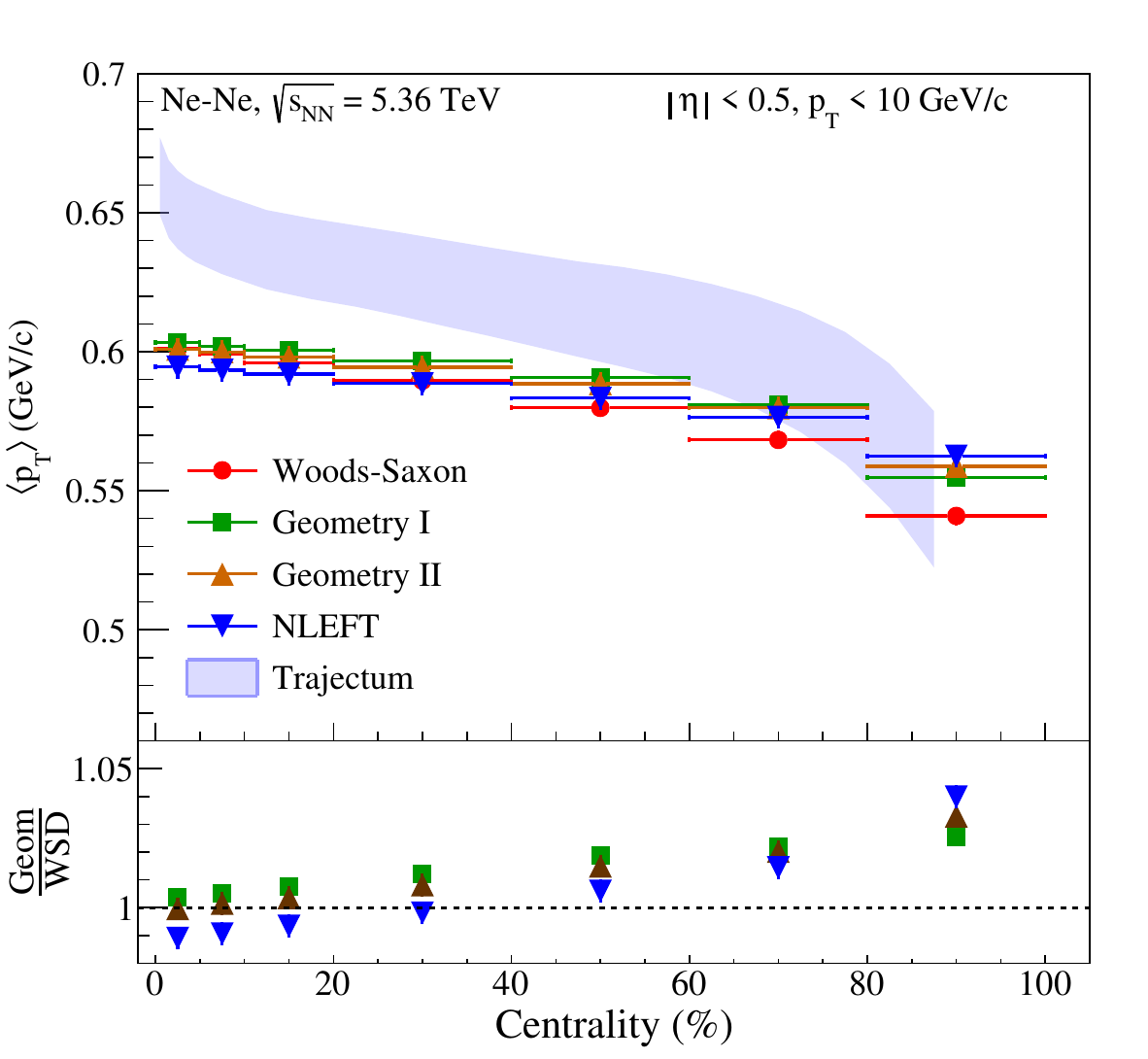}% Here is how to import EPS art
\caption{\label{fig:8} (Top panel) Mean transverse momentum ($\langle p_T \rangle$) of charged hadrons at mid-rapidity in $^{20}$Ne$-$$^{20}$Ne collisions at $\sqrt{s_{NN}}$ = 5.36 TeV for Woods-Saxon, NLEFT and bi-pyramidal geometries as a function centrality classes. Trajectum model results are also shown as solid band. Bottom panels present the ratios of NLEFT and Geometry I and II to the Woods–Saxon spectra.}    
\end{figure}

Fig.~\ref{fig:8} shows the mean transverse momentum ($\langle p_T \rangle$) of charged hadrons as a function of centrality for the Woods-Saxon and bi-pyramidal geometries of the $^{20}$Ne nucleus. For comparison, results of the Trajectum hydrodynamical model are also presented. Bottom panel represents the $\langle p_T \rangle$ ratios of bi-pyramidal geometries to the Woods-Saxon distribution. The mean transverse momentum, $\langle p_T \rangle$, is observed to decrease with increasing centrality of the collisions for all the considered configurations. Geometry I and Geometry II exhibit nearly identical $\langle p_T \rangle$ values across all centrality classes, indicating negligible sensitivity to the specific bi-pyramidal geometry. The NLEFT configuration shows a slightly lower $\langle p_T \rangle$ in central collisions and toward peripheral collisions it exhibits the highest $\langle p_T \rangle$ among all the configurations. Although the differences among the bi-pyramidal configurations remain small in central collisions, a growing separation between the bi-pyramidal geometries and the Woods–Saxon distribution is observed as the collisions become more peripheral. In the peripheral collisions, the Woods–Saxon distribution yields the lowest $\langle p_T \rangle$, as also evident from the Geometry to Woods–Saxon ratios shown in the lower panel. This behaviour of $\langle p_T \rangle$ can be explained within the Angantyr model on the basis of multi-parton interactions and the resulting string densities based on the number of effective nucleon-nucleon collisions. The nearly negligible variation of $\langle p_T \rangle$ in central collisions arises primarily from the large overlap volume where geometric substructures are effectively averaged out, making $\langle p_T \rangle$ insensitive to differences in nuclear geometry. In contrast, in mid-central and peripheral collisions, the reduced overlap region enhances the sensitivity to the spatial $\alpha$-cluster distributions which leads to the geometry driven variations in the number of effective hard and semi-hard scatterings. \par

On comparison of the Angantyr configurations with the Trajectum model results, it is observed that the $\langle p_T \rangle$ predicted by Trajectum is significantly higher than that obtained from Angantyr in central collisions. As the collisions become non-central, this difference gradually decrease, and the $\langle p_T \rangle$ values Angantyr and Trajectum tend to converge toward peripheral collisions. The observed behaviour is due to the presence of collective hydrodynamic expansion in Trajectum model which generates additional radial flow and pushes particles to higher transverse momentum in central, high density collisions. In contrast, Angantyr does not include hydrodynamic evolution and relies on independent nucleon-nucleon interactions and string fragmentation which lead to a comparatively softer $\langle p_T \rangle$. As the system becomes smaller and less dense in peripheral collisions, the collective flow effects are progressively reduced in Trajectum resulting in a convergence of $\langle p_T \rangle$ with that obtained from the Angantyr model.\par

\begin{figure}[h]
\includegraphics[width=\columnwidth]{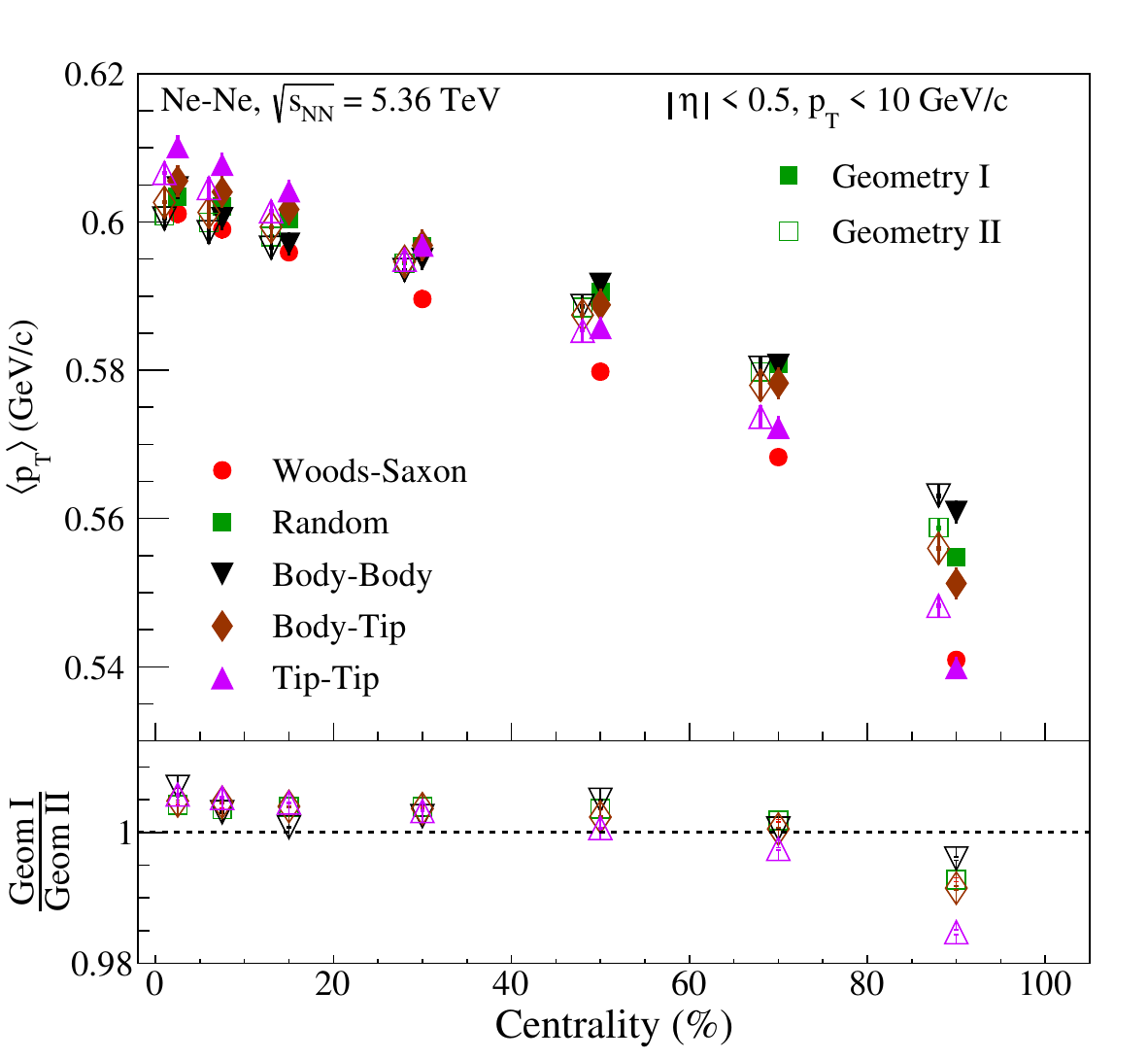}% Here is how to import EPS art
\caption{\label{fig:9} (Top panel) Mean transverse momentum ($\langle p_T \rangle$) of charged hadrons at mid-rapidity in $^{20}$Ne$-$$^{20}$Ne collisions at $\sqrt{s_{NN}}$ = 5.36 TeV for Geometry I and Geometry II in Body-Body, Body-Tip and Tip-Tip orientations of $^{20}$Ne nucleus. For better visibility, the Geometry II results are shown with open markers and are slightly shifted to the left. Bottom panel presents the Geometry I to Geometry II ratio for the corresponding nuclear orientations.}    
\end{figure}

Fig.~\ref{fig:9} shows the variation of $\langle p_T \rangle$ of charged hadrons in $^{20}$Ne-$^{20}$Ne collisions at $\sqrt{s_{NN}}$ = 5.36 TeV for Geometry I and Geometry II in BB, BT, TT and random orientations of $^{20}$Ne nucleus along with Woods-Saxon distribution. The Geometry I to Geometry II ratio for the corresponding nuclear orientations are also presented in bottom panel. In central collisions, the $\langle p_T \rangle$ is highest for TT orientation for both Geometry I and Geometry II. The BB, BT, and random orientations yield nearly identical $\langle p_T \rangle$ values, with only marginal differences within the uncertainties for both the geometries. The enhanced $\langle p_T \rangle$ in TT collisions arises from the larger local overlap density and higher string density associated with the alignment of the nuclear long axes, leading to an increased contribution from multi-parton interactions. In contrast, BB, BT, and random orientations involve large overlap volumes where geometric substructures are effectively averaged out, resulting in similar effective interaction densities and hence nearly identical $\langle p_T \rangle$ values.\par

As the collisions become non-central, the mean transverse momentum in TT collisions gradually decreases and approaches that of the random orientation. In the 20–40\% centrality interval, the TT results overlap with the random configuration, and for more peripheral collisions, the TT $\langle p_T \rangle$ becomes lower than that of the random, body–body, and body–tip orientations for both Geometry I and Geometry II. This behaviour is due to the progressive reduction in the overlap density of TT collisions as the impact parameter increases. In non-central collisions, the alignment advantage of the long axes is diminished, reducing the effective string density and the number of semi-hard scatterings. Consequently, the $\langle p_T \rangle$ in TT collisions becomes comparable to and eventually smaller than random and other orientations in peripheral collisions where fluctuations in nucleon and cluster overlap continue to contribute effectively.
In the same centrality range, the BT configuration also falls below the random orientation but remains higher than the TT configuration for both Geometry I and Geometry II. The BT configuration, while still benefiting from partial alignment, experiences a similar but less pronounced reduction. The BB orientation exhibits a non-monotonic behavior. As the collisions become non-central, the separation between its mean transverse momentum and that of the other orientations initially increases, making BB the lowest in $\langle p_T \rangle$. In the 20–40\% centrality interval, the BB results overlap with those of all other orientations. Beyond this centrality, the BB configuration yields the highest $\langle p_T \rangle$ among all orientations in the peripheral regime for both Geometry I and Geometry II. This behavior can be understood in terms of the transverse spatial distribution of $\alpha$-clusters in the BB configuration. In mid-central collisions, the broad transverse extent of BB geometry leads to a reduction in local overlap density, suppressing the effective number of semi-hard scatterings and resulting in a lower $\langle p_T \rangle$. However, in peripheral collisions, the extended transverse distribution increases the likelihood that a subset of $\alpha$-clusters remains within the interaction zone, enhancing local string densities and multi-parton interactions relative to TT, BT, and random  orientations. Consequently, the BB orientation produces the largest $\langle p_T \rangle$ in the peripheral region.\par

The observation of distinct $\langle p_T \rangle$ values for different collision configurations is basically due to the variation in the overlap geometry of high and low density regions arising from $\alpha$-clustered structure of the $^{20}$Ne and smooth nucleonic density distribution for Woods-Saxon geometry.

\section{Conclusions}
In conclusion, we have investigated the effect of the intrinsic structure of the $^{20}$Ne nucleus on the dynamics of small collision systems using the Pythia8/Angantyr model. By implementing the bi-pyramidal $\alpha$-cluster structure of $^{20}$Ne in two distinct configurations along with NLEFT configurations and Woods-Saxon distribution, we have explored the influence of the elongated structure of $^{20}$Ne on the particle production patterns in $^{20}$Ne$-$$^{20}$Ne collisions at $\sqrt{s_{NN}}$ = 5.36 TeV. We find that charged-particle multiplicity observables are notably sensitive to the underlying nuclear geometry in the Angantyr model. Configurations with explicit $\alpha$-cluster structure exhibit clear deviations from the Woods–Saxon results, particularly in central and mid-central collisions, reflecting differences in the geometric overlap region and the number of effective nucleon–nucleon interactions. Orientation dependent studies further reveal that Tip–Tip, Body-Body, and Body–Tip orientations lead to distinct multiplicity patterns, with these differences diminishing toward peripheral collisions as the overlap region becomes smaller and less sensitive to the details of the nuclear structure. A higher mean multiplicity is observed for Tip-Tip, Body-Body, and Body-Tip orientations in Geometry I compared to Geometry II. It is primarily due to the smaller radius of the colliding nuclei across different orientations in Geometry I, which leads to a higher string density and consequently a larger average multiplicity. In general, these patterns depend on how a given initial state model converts the nuclear geometry and nuclear orientation into string density (Angantyr) or entropy (Trajectum) production in the early stage of the collision. As a result, the relative multiplicities for these orientations can vary depending on the initial nuclear geometry in different models. \par

 In contrast, the transverse momentum distributions and mean transverse momentum show only modest sensitivity to nuclear geometry and clustering. While small orientation and geometry dependent variations in $\langle p_T \rangle$ are observed in non-central collisions, the overall behavior is largely governed by the number of effective nucleon–nucleon interactions and multi-parton interactions inherent to the Angantyr framework arising from the spatial distribution of $\alpha$-clusters. Comparisons with hydrodynamic model results indicate that collective expansion effects, absent in Angantyr, play a dominant role in shaping transverse momentum observables in central collisions.\par

The findings highlight the sensitivity of small system dynamics to the underlying nuclear geometry, demonstrating that the intrinsic $\alpha$-cluster structure of light nuclei such as $^{20}$Ne leaves observable imprints on particle production, even in the absence of hydrodynamic evolution. The present results provide a systematic baseline for disentangling initial state geometric effects from medium induced dynamics in small collision systems. It would be of further interest to investigate the role of different geometrical configurations and orientations of $^{20}$Ne nucleus on anisotropic flow observables within the Angantyr framework so that forthcoming experimental results can be compared with and without medium effect scenarios, thereby offering valuable insight into the interplay between nuclear structure and collective dynamics.

\begin{acknowledgments}
The authors thank Giuliano Giacalone and Wilke van der Schee for kindly providing the Trajectum model results. S.D. and A.S. acknowledge the SERB Power Fellowship, No. SPF/2022/000014, for the support on this work.
\end{acknowledgments}

\appendix

\nocite{*}

\bibliography{manuscript}% Produces the bibliography via BibTeX.

\end{document}